# Engineering of Form Factor in Rotationally Symmetric Photonic Cavities for Faster Axion Searches

Mohamed H. Awida

*Abstract*—Form factor can be engineered in photonic cavities such that it dramatically increases the scan speed of axion haloscope detectors. Using higher-order modes of a photonic cavity, we can probe axions at larger mass ranges beyond the reach of current detectors that rely on conventional copper cavities. The main idea is to produce a higher-order mode with longitudinal field distribution that has relatively high amplitude away from both the cavity axis and the cavity walls such that it results in a relatively large form factor and maintains a high-quality factor. The detector scan speed significantly increases as it is proportional to the square of the form factor multiplied by the quality factor.

## I. Introduction

HALOSCOPE detectors rely on resonant cavity structures residing in a uniform magnetic field to detect axions through the inverse Primakoff effect [1]. Not only do the axions solve the strong-CP problem [2], but they are also a primary contender for dark matter detection. This double motivation prompted many experiments all over the world to intensify effort to probe axions at different mass ranges; ADMX in the mass range 2-4µeV [3]-[4], CAPP at 6.7µeV [5], HAYSTACK at 24µeV [6], QUAX at 43µeV [7], ORGAN at 110µeV [8]. Haloscopes have the advantage of signal enhancement by forming standing waves at the cavity's resonant frequencies [1]. Standard pillbox copper cavities are conventionally used in these haloscopes as they feature a large volume of interaction between the electromagnetic field inside the cavity and the magnetic field. Haloscopes typically employ solenoid magnets [3]-[8]. The cavity and magnet interaction is quantified in the form factor parameter [9].

Different models exist to express the expected axion coupling. Among them, two plausible models are prevalent to determine the sensitivity limit a detector should aim for, namely, KSVZ [10]-[11] and DFSZ [12]-[13]. A single cavity was successfully used by the ADMX collaboration for axion detection to reach the DFSZ limit [3]-[4]. However, to scan at larger mass ranges, the physical cavity dimensions will scale unfavorably down with frequency. This scale down with mass range (frequency) necessitates the usage of multiple cavities to maintain the detection volume, which in turn determines the haloscope sensitivity and the scan rate. However, working with multiple cavities adds significant complexity to the detector because of the need to coherently combine the output signal of the multiple cavities.

On the other hand, photonic cavities represent another path for haloscope detectors at relatively high mass ranges [14]-[17]. Loading the cavity with dielectrics opens a path forward for using a higher-order mode other than TM010, enabling a relatively large detection volume for the cavity while operating at a higher frequency range. Meanwhile, if it is engineered properly, dielectric loading can produce ultra-high quality factor cavities beyond the reach of copper cavities [17]-[23].

In the microwave community, it is common to refer to those dielectric-loaded cavities as distributed Bragg reflectors or resonators (DBR) [18]-[23]. There was significant work on such structures, whether in cylindrical [18]-[22] or spherical geometries [23] using transverse electric modes for microwave applications. For instance, the unloaded quality factor of 7e5 was demonstrated at room temperature, however, with a transverse electric (TE) mode rather than a transverse magnetic (TM) one [19].

Recently, the QUAX collaboration demonstrated a photonic cavity design intended for axion searches: a cylindrical copper cavity loaded with two hollow dielectric shells, which is basically a DBR but utilizes a transverse magnetic mode that is needed for axion searches to get a plausible form factor. The cavity operated with the TM030 mode, exhibiting a measured unloaded quality factor of 7.2e5 (cold) but with a modest form factor of 0.04 [17]. This development opens many possibilities for future high Q cavities to be used in haloscope detectors, especially if combined with quantum sensors that act as photon counters to evade the quantum limit and relax the requirements on the detection volume [24]. For such photon counters, the high Q cavity is needed to allow the quantum sensors employed in the photon counter to repeatedly do enough parity measurements for noise suppression [25].

It is worth mentioning here that there is also significant work on the direction of superconducting high Q cavities for axion searches. The grand challenge for a superconducting cavity is to maintain a high-quality factor in a high magnetic field. Promising results were obtained with NbTi [26] and YBCO [27]-[28]. However, photonic cavities promise a relatively larger detection volume by utilizing a higher-order mode of the cavity structure.

This work was supported by the U.S. Department of Energy under Contract Number DE-AC02-07CH11359.

Author is with the Fermi National Laboratory, Batavia, IL 60510 USA (630-840-3935; e-mail: mhassan@fnal.gov).

On the other hand, a major hurdle for photonic cavities is maintaining their performance in terms of form and quality factors while achieving a significant tuning range. Tuning a photonic cavity is rather more difficult than a standard pillbox copper cavity as a metallic rod cannot generally be employed in the high field regime. In this case, the RF losses on the metallic rod itself will degrade the cavity's quality factor. Some work on photonic cavities for axions addressed the tunability challenge by integrating the tunning mechanism in the dielectric loading itself. Two methods were used either by utilizing supermodes [14] or through dielectric wedges and working with an azimuthally varying transverse magnetic mode [16]. However, both the supermodes method and the wedge implementation ended up so far with cavities limited in quality factor by the wall losses as the field distribution is relatively high near the cavity walls. On the other hand, the QUAX collaboration suggested a conceptual mechanism for tuning rotationally symmetric cavities by dividing the inner shell longitudinally into two halves and moving them against each other in the transverse plan [17].

In this letter, we discuss a potential technique to engineer rotationally symmetric photonic cavities such that they exhibit a high form factor >0.1 while maintaining a high-quality factor >5e5, beyond state of the art in photonic cavities for axion searches. The superior performance of the proposed cavities immediately reflects on the scan rate as it is quadratically proportional to the form factor and is linearly proportional to the quality factor [9].

## II. FORM FACTOR ENGINEERING

Conventionally, the form factor of a resonant mode inside the RF cavity is defined as [9]

$$C_{mnp} = \frac{|\int dx^3 B_0 \cdot E_{mnp}(x)|^2}{B_0^2 V \int dx^3 \varepsilon(x) |E_{mnp}(x)|^2}$$

Where $B_0$ is the dc magnetic field in the interaction region of the magnet and the cavity, $E_{mnp}$ is the electric field of an electromagnetic mode number $mnp$ in the cavity, V is the cavity volume, and $\varepsilon(x)$ is the dielectric constant as a function of volume. Both integrals in the form factor definition are carried over the cavity volume.

The most common magnet type in axion searches is a solenoid, where the magnetic field is vertical. The form factor, in this case, can then be reduced to

$$C_{mnp} = \frac{|\int dx^3 Ez_{mnp}(x)|^2}{V \int dx^3 \varepsilon(x) |E_{mnp}(x)|^2}$$

where $Ez$ is the electric field in the vertical direction. The solenoid magnet case narrows down the useful resonant modes to TM modes.

Moreover, the highest in form factor among the TM modes will be the one with the full field in the z-direction; TM0n0 (assuming a mode convention of a cylindrical cavity). There are no longitudinal variations to the field for such modes, and the electromagnetic problem can then be simplified to 2D to study only the cavity's cross-section.

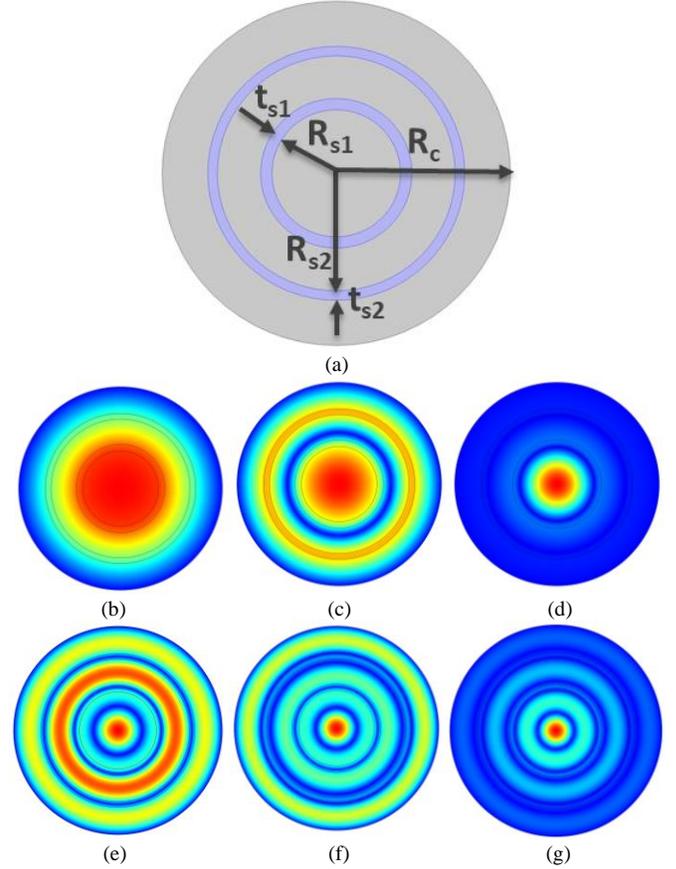

Fig. 1. Electromagnetic resonance modes in photonic cavity with two dielectric shells. (a) 2D cross-section of the cavity. (b) TM010. (c) TM020. (d) TM030. (e) TM040. (f) TM050. (g) TM060.

Let's consider a photonic bandgap cavity loaded with two layers of nested sapphire tubes as shown in Fig. 1(a), similar to the work presented before by QUAX in [17]. Figure1(b)-(g) show the simulated longitudinal electric field for the first six TM0n0 modes. Comsol Multiphysics was used in these electromagnetic simulations [29]. We have intentionally chosen the dimensions of $R_c$=29 mm, $R_{s1}$=10.7 mm, $t_{s1}$=1.9 mm, $R_{s2}$=19.8 mm, and $t_{s2}$=1.6 mm, that are identical to the cavity in [17] to use it as a reference for comparison purposes. Dielectric constant of 11.2 and loss tangent of 2e-6 was assumed for the sapphire.

Meanwhile, Table 1 lists the resonance frequencies, form, and quality factors for these six TM0n0 modes. The table also lists $C^2Q_0$, which is a direct measure of the scan rate. TM030 is the mode that was used in [17], where the simulated form factor in 2D is 0.041, and the quality is 2.33e6. The work in [17] focused on increasing the cavity's quality factor to the maximum extent. In that sense, TM030 is the highest Q mode, as shown in Table 1. We noticed, however, while studying the higher-order modes of this structure that some modes in the same cavity exhibit superior form factors, specifically, TM050 and TM060. TM060, in particular, also exhibits a relatively high-quality factor resulting in the highest $C^2Q_0$, as listed in Table 1.

The superior performance of TM060 warranted further investigation to come up with a physical explanation as to why this mode exhibits such properties. This performance, to our knowledge, was not reported before in the literature.

Table 1. List of the resonance frequency, form and quality factors, and $C^2Q_0$ for the first six TM0n0 resonance modes in a photonic bandgap cavity with two shells of dielectrics.

| Mode | Freq [GHz] | C | $Q_0$ | $C^2Q_0$ |
|---|---|---|---|---|
| TM010 | 2.389 | 0.251 | 1.07E+05 | 6707 |
| TM020 | 5.631 | 0.027 | 1.63E+05 | 122 |
| TM030 | 10.883 | 0.041 | 2.33E+06 | 3908 |
| TM040 | 16.932 | 0.002 | 2.18E+05 | 0.8 |
| TM050 | 19.574 | 0.268 | 1.46E+05 | 10515 |
| TM060 | 22.857 | 0.166 | 6.05E+05 | 16693 |

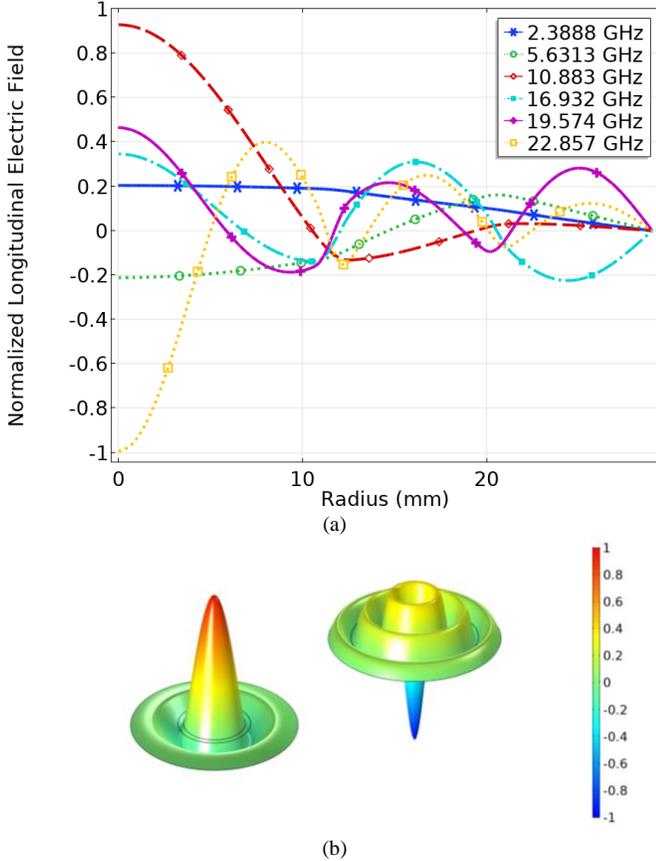

Fig. 2. Longitudinal electric field across the PBG cavity. (a) Normalized electric fields for the first six TM0n0 modes versus the cavity radius. (b) Comparison of the electric fields for TM030 (10.88 GHz) on the left and TM060 (22.857 GHz) on the right.

Figure 2(a) depicts the simulated longitudinal electric field for the first six TM0n0 modes normalized to the same energy level. TM030 is in red, and it does have the three expected oscillation in the radial direction, while TM060 is in yellow and have six oscillations. A first look at the graph might deceive the reader that the TM060 will cancel out in form factor integrals. However, the form factor has volume integrals, not line integrals, which essentially means that the fields are weighted by their radial distance from the cavity axis, giving much larger weight to the fields when they are further away from the axis.

The conventional way of cavity engineering for better form factor used to be in securing large field content in the first oscillation of the field near the cavity axis and minimizing the field distribution afterward. TM010 is the best mode in this perspective and exhibits the highest form factor, and that is what most conventional haloscopes are using for their empty copper cavities. However, with the dielectric loading in photonic bandgap cavities, there is room to engineer the form factor and use a higher-order mode beyond TM010.

Meanwhile, TM030 was also a choice for some photonic cavities, mainly because it has a shallow electric field near the copper wall, as shown in Figures 2(a) and (b). Therefore, it exhibits a superior quality factor. However, that mode doesn't exhibit a large form factor because it again relies only on the first oscillation of the electric field to dominate other oscillations, which is hard to achieve given the larger weight for the off-axis field content. On the other hand, a paradigm shift in photonic cavities will be to engineer the cavities such that they produce a mode with larger content on the outward oscillations further away from the cavity axis—this way of form factor engineering results in a relatively high form factor. Meanwhile, keeping a significant decrease in the fields towards the copper walls secures a decently high quality factor, similar to what we observed for TM060, as shown in Figures 2(a) and(b).

To prove that this can be generalized, let us consider a simpler cavity case with one dielectric shell, as shown in Fig. 3(a). We kept the cavity radius the same as the case of two dielectric shells; Rc=29 mm and optimized the dielectric shell dimensions to get superior performance for TM040. We ended up with the following dimensions: Rs=16.5 mm, and ts=1.7mm. Figure 3(b)-(e) show the simulated longitudinal electric field for the first four TM0n0 resonant modes in the cavity structure. The RF module of Comsol Multiphysics was again used in these simulations. Table 2 lists the parameters of these four resonance TM0n0 modes in terms of frequency, form, and quality factors, and $C^2Q_0$. As detailed in the table, we managed to get a form factor of 0.143 and a quality factor of 5.7e5 for TM040 at 16.081 GHz. This simultaneous good performance in both form and quality factors potentially enables an axion search at this high mass range.

Without loss of generality, the proposed methodology of form factor engineering can be translated to other photonic cavity structures at different frequency ranges. The implications of the proposed method for form factor engineering will be discussed in the next section in terms of the challenges of mode crowding and tunning.

### III. DISCUSSION

The proposed method of form factor engineering represents a paradigm shift from localizing the electric field on the cavity axis to distributing it more off-axis, utilizing more of the cavity volume further away from the axis. It is viable in producing higher-order modes with superior performance in terms of form and quality factors. However, two issues arise here, namely, mode crowding and tunability. On the one hand, the usage of higher-order mode faces the difficulty of mode crowding, and the issue becomes more pronounced with getting higher in the mode order number. For instance, mode crowding in the case of TM060 will be an issue that may prevent utilizing that mode efficiently in a haloscope, while it seems more manageable for TM030 and TM040. Nevertheless, this opens a window of research opportunity to investigate means to inhibit the parasitic modes around the useful mode.

On the other hand, the tunability of photonic bandgap cavities, especially the kind discussed in this paper that keeps rotational symmetry, remains an open question that lacks a physical demonstration. It is worth noting here that tuning a higher-order mode will be more challenging as the mode order number increases. The method of tuning suggested by the QUAX collaboration in [17] of moving two half-shells in the transverse plan might be feasible for tuning the class of cavities discussed in this paper.

## IV. CONCLUSION

Engineering of form factor in rotationally symmetric photonic bandgap cavities is possible to produce superior modes of simultaneously high form and quality factors. Electromagnetic higher-order modes with field content farther away from the axis exhibit high form factor and can be optimized to exhibit low field near the cavity walls to produce also a high-quality factor. We have demonstrated that TM060 in a two-shell dielectric-loaded photonic cavity can potentially have C of 0.16 and $Q_0$ of 6e5. Similarly, we have also shown that TM040 in a single shell dielectric-loaded cavity can exhibit C of 0.14 and $Q_0$ of 5e5. We advocate for a paradigm shift to better engineer photonic cavities for haloscope detectors that rely on the proposed form factor engineering method. We have also discussed the implications of the proposed methodology in terms of mode crowding and tunability.

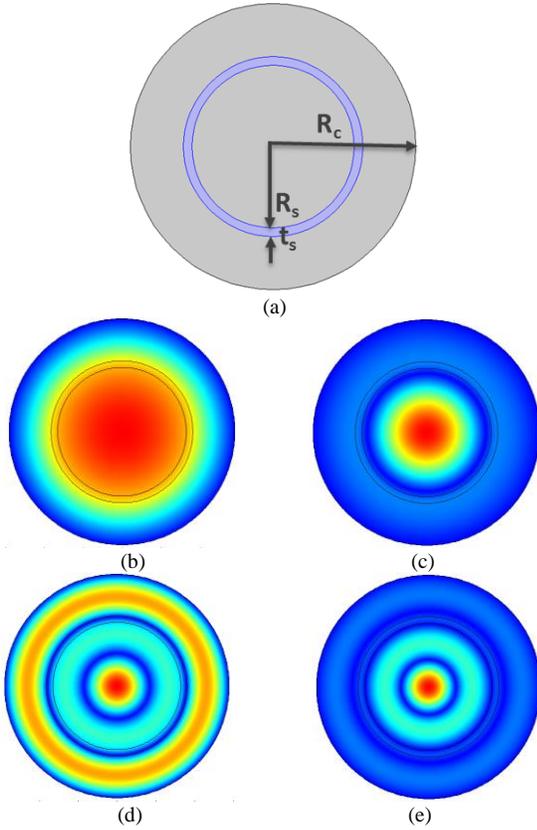

Fig. 3. Electromagnetic resonance modes in photonic bandgap cavity with one dielectric shell. (a) 2D cross-section of the cavity. (b) TM010. (c) TM020. (d) TM030. (e) TM040.

Table 2. List of the resonance frequency, form and quality factors, and $C^2Q_0$ for the first four TM0n0 resonance modes in a photonic bandgap cavity with one dielectric shell.

| Mode  | Freq [GHz] | C      | $Q_0$    | $C^2Q_0$ |
|-------|------------|--------|----------|----------|
| TM010 | 2.862      | 0.3579 | 1.04E+05 | 13329    |
| TM020 | 7.560      | 0.0214 | 7.18E+05 | 328      |
| TM030 | 13.457     | 0.2631 | 1.24E+05 | 8552     |
| TM040 | 16.081     | 0.1433 | 5.73E+05 | 11772    |

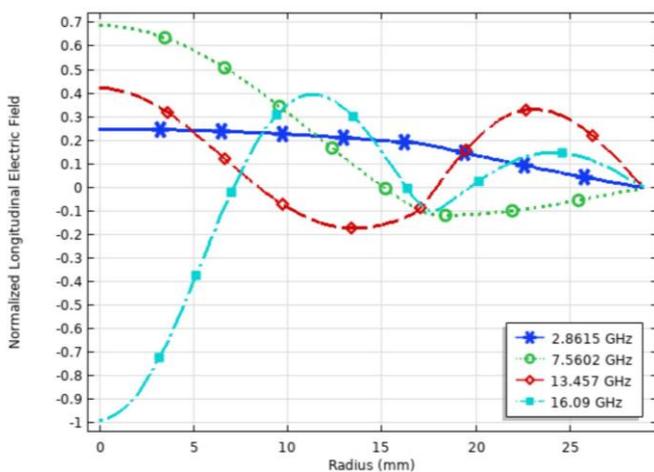

Fig. 4. Longitudinal normalized electric field across the PBG cavity with one dielectric shell for the first four TM0n0.